\documentclass[%
aip,
jmp,
 amsmath,amssymb,
reprint,%
]{revtex4-2}

\usepackage{graphicx}% Include figure files
\usepackage{dcolumn}% Align table columns on decimal point
\usepackage{bm}% bold math
\usepackage[utf8]{inputenc}
\usepackage[T1]{fontenc}
\usepackage{mathptmx}
\usepackage{color}
\usepackage{etoolbox}
\usepackage{braket}
\usepackage{amsmath}
\usepackage{dsfont}
\usepackage{hyperref}
\usepackage{comment}
\allowdisplaybreaks

\newcommand{\Ber}{\text{Ber}}
\renewcommand{\Im}{\text{Im}}
\newcommand{\sgn}{\text{sgn}}
\newcommand{\tr}{\text{tr}\,}
\newcommand{\Pf}{{\text{Pf}}}
\newcommand{\abs}[1]{\ensuremath{\left\vert#1\right\vert}}

\makeatletter
\def\@email#1#2{%
 \endgroup
 \patchcmd{\titleblock@produce}
  {\frontmatter@RRAPformat}
  {\frontmatter@RRAPformat{\produce@RRAP{*#1\href{mailto:#2}{#2}}}\frontmatter@RRAPformat}
  {}{}
}%
\makeatother
\begin{document}

\preprint{AIP/123-QED}

\title[Winding Number Statistics for Chiral Random Matrices]{Winding Number Statistics for Chiral Random Matrices: Averaging Ratios of Parametric Determinants in the Orthogonal Case}

\author{Nico Hahn$^{1*}$, Mario Kieburg$^2$, Omri Gat$^3$ and Thomas Guhr$^1$}
\affiliation{$^1$ Fakult\"at f\"ur Physik, Universit\"at Duisburg--Essen, Duisburg, Germany\\
             $^2$ School of Mathematics and Statistics, The University of Melbourne, Melbourne, Australia\\
             $^3$ Racah Institute of Physics, The Hebrew University, Jerusalem, Israel
             }
\email{nico.hahn@uni-due.de}

\date{\today}% It is always \today, today,

\begin{abstract}
We extend our recent study of winding number density statistics in Gaussian random matrix ensembles of the chiral unitary (AIII) and chiral symplectic (CII) classes. Here, we consider the chiral orthogonal (BDI) case which is the mathematically most demanding one. The key observation is that we can map the topological problem on
a spectral one, rendering the toolbox of random matrix theory applicable. In particular, we employ a technique that exploits supersymmetry structures without reformulating the problem in superspace.
\end{abstract}

\maketitle

\section{Introduction}
\label{secI}

The topological classification of spectrally gapped Hamiltonians plays a major role in contemporary condensed matter physics. This is due to the possible emergence of localized states at the boundary of the system, which arrive with a certain robustness against perturbations \cite{BH2011, OLZH2022, CFYRSB2022, CXYKT2021, BWPE2019}. The number of edge states is related to a topological invariant that is defined for the closed system with periodic boundary conditions, referred to as the bulk. This relation is known as bulk-boundary correspondence and has been proven recently to have far-reaching validity \cite{ProdanBook, Alldridge2020}.

The symmetries of the system play an important role for the existence of a non-trivial topology, and hence of edge states. The presence of disorder, i.e. spatially inhomogeneous perturbations of the system parameters, excludes spatial symmetries, such as rotations. There are ten remaining symmetry classes, summarized in the tenfold way classification \cite{AltlandZirnbauer1997, Heinzner2005}. In its framework, it can be predicted whether a non-trivial topology is possible in dependence of the symmetry class and the dimension of the system \cite{Schnyder2008, Kitaev2009}, see Ref.~\onlinecite{Ryu2016} for an overview.

Chiral symmetry is one of the symmetries of the tenfold way classification. It finds its origin in quantum chromodynamics, where the chiral symmetry of the Dirac operator is broken by the quark masses and is restored in the high temperature limit. In condensed matter systems it appears either as a combination of time reversal invariance and a particle-hole constraint \cite{Zirnbauer2021} or as a sublattice symmetry \cite{Gade1993, AsbothBook}. Assuming chiral symmetry, one can further distinguish between three different chiral symmetry classes based on time reversal invariance: the chiral unitary class with broken time reversal invariance, and the time reversal invariant chiral orthogonal and chiral symplectic classes (labeled AIII, BDI and CII, respectively, in the tenfold way).

In one-dimensional chiral symmetric systems the pertinent topological invariant is the winding number \cite{AsbothBook, Maffei2018}. In general, the topological
invariant is sensitive to the disorder configuration, which motivates a statistical
description in which the invariant becomes a random variable. We simulate disorder by setting up a parametric random matrix model for the Hamiltonian, in which we want to analyze the winding number statistics. Random matrix theory is known to successfully model universal properties in systems with a sufficient degree of complexity in the limit of large matrix dimensions \cite{MehtaBook, GuhrRMTReview}. Our work is preceded by the fruitful application of chiral random matrices in quantum chromodynamics \cite{VerbaarschotZahed1993, ShuryakVerbaarschot1993, Verbaarschot1994, Wettig1996A, Wettig1996B, JacksonVerbaarschot1996, VerbaarschotWettig2000, GWW2000} and of parametric random matrices in disordered systems \cite{SimonsAltshuler1993A, SimonsAltshuler1993B, BeenakkerRejaei1994}. But unlike these earlier results using similar random matrix models, we are not investigating spectral properties, but topological ones. This paper ties in directly with our previous works, where we analyzed the statistics of the winding number in the chiral classes AIII and CII \cite{BHWGG2021, HKGG2023}. Here, we want to catch up on the chiral orthogonal class BDI.

Our goal is to calculate averages over ratios of determinants with parametric dependence. These will serve as generators for the winding number statistics. The problem is related to averaging ratios of characteristic polynomials, which is a common task in random matrix theory \cite{FyodorovStrahov2003, BorodinStrahov2006, KieburgGuhr2010a, KieburgGuhr2010b, ASW2020, IF2018, Webb2015, MN2001, CFKRS2003, BHNY2007, Eberh2022, Fyod2004}, because they entail information about the spectral statistics. Correspondingly, powerful methods, like the supersymmetry method \cite{Efetov1983}, have been developed to treat such ensemble averages. In this way, we are mapping a topological problem to a spectral one.

In this study, just as we did in Ref.~\onlinecite{HKGG2023}, we employ a technique introduced by two of the present authors \cite{KieburgGuhr2010a, KieburgGuhr2010b}. It establishes exact expressions for the ensemble averages while also revealing supersymmetric structures without actually mapping to superspace. For this reason it has been coined supersymmetry without supersymmetry. The resulting expressions are determinants in the unitary case and Pfaffians in the orthogonal and symplectic case containing simplified averages of only two determinants. The remaining task is to calculate these simplified averages over the corresponding random matrix ensemble, which is the step we show in this paper.

The outline of the paper is as follows: in Sec.~\ref{secII} we set up the random matrix model and define the problem we want to solve. In Sec.~\ref{secIII} we present our results, while Sec.~\ref{secIV} contains the derivations. We relegated some details to the appendix. In Sec.~\ref{secV} we conclude our study.

\section{Posing the Problem}
\label{secII}

Formally, chiral symmetry of a Hamiltonian $H$ can be expressed as
\begin{equation} \label{2ChiralSymmetry}
\{\mathcal{C},H\} = 0\quad {\rm with}\quad \mathcal{C}^2=\mathds{1},
\end{equation}
where $\{,\}$ denotes the anticommutator and $\mathcal{C}$ is the chiral operator. In its eigenbasis it reads as
\begin{equation}
\mathcal{C} = \begin{bmatrix}
\mathds{1} & 0
\\
0 & -\mathds{1}
\end{bmatrix}
\end{equation}
and the Hamiltonian takes a block off-diagonal form
\begin{equation} \label{2ChiralHamiltonian}
H = \begin{bmatrix}
0 & K
\\
K^\dagger & 0
\end{bmatrix},
\end{equation}
where the matrix $K$ and its Hermitian adjoint $K^\dagger$ have dimension $N\times N$. We want to consider the chiral orthogonal symmetry class, labeled BDI in the tenfold way. In this class the time reversal operator $\mathcal{T}$ is given by a complex conjugation, such that $\mathcal{T}^2 = +1$, and the Hamiltonian fulfills the equation
\begin{equation}
\mathcal{T} H \mathcal{T}^{-1} = H^* = H,
\end{equation}
with $(\cdot)^*$ being the complex conjugation. This renders $H$ and $K$ real symmetric and real, respectively. In random matrix theory this case is usually labeled as $\beta = 1$, where the Dyson index $\beta$ is the dimension of the underlying number field. Correspondingly, class AIII, featuring complex entries and class CII, featuring quaternion entries are labeled by $\beta = 2$ and $\beta = 4$, respectively.

In the following we choose the dimension $N$ of the off-diagonal blocks as even. This is due to technical reasons, that exist exclusively for the orthogonal case. However, this does not interfere with our goal to uncover universal behaviour in the large $N$ limit as is outlined in more detail further below. 

Drawing all independent elements of $H$ from a Gaussian probability distribution leads to the chiral Gaussian orthogonal ensemble (chGOE), while at the same time $K$ can be viewed as a member of the real Ginibre ensemble \cite{Ginibre1965}. The matrix probability densities of these ensembles are invariant under orthogonal transformations.

To examine topological properties we endow the Hamiltonian with a parametric dependence, $H = H(p)$ and thus $K = K(p)$, that we assume to be periodic $H(p+2\pi) = H(p)$. Physically, the parametric dependence emerges naturally when $H(p)$ is interpreted as a Bloch Hamiltonian that is obtained as the Fourier transform from real space to momentum space \cite{Ryu2016, BernevigBook}. The parameter $p$ is then referred to as a (quasi-)momentum taking its values in the one-dimensional Brillouin zone $[0,2\pi)$. Upon time reversal the parametric dependence behaves as
\begin{equation}	\label{2parametricTRI}
\mathcal{T}K(p)\mathcal{T}^{-1} = K^*(p) = K(-p).
\end{equation}
Consequently, $K(p)$ will be complex in general and real only at the time reversal invariant momenta $p=0$ and $p=\pi$, where $p$ corresponds to $-p$ due to the $2\pi$-periodicity. Then also $H(p)$ will be non-real, despite being in the class BDI.

\begin{figure}[t!]
\centering
\includegraphics[width=\textwidth]{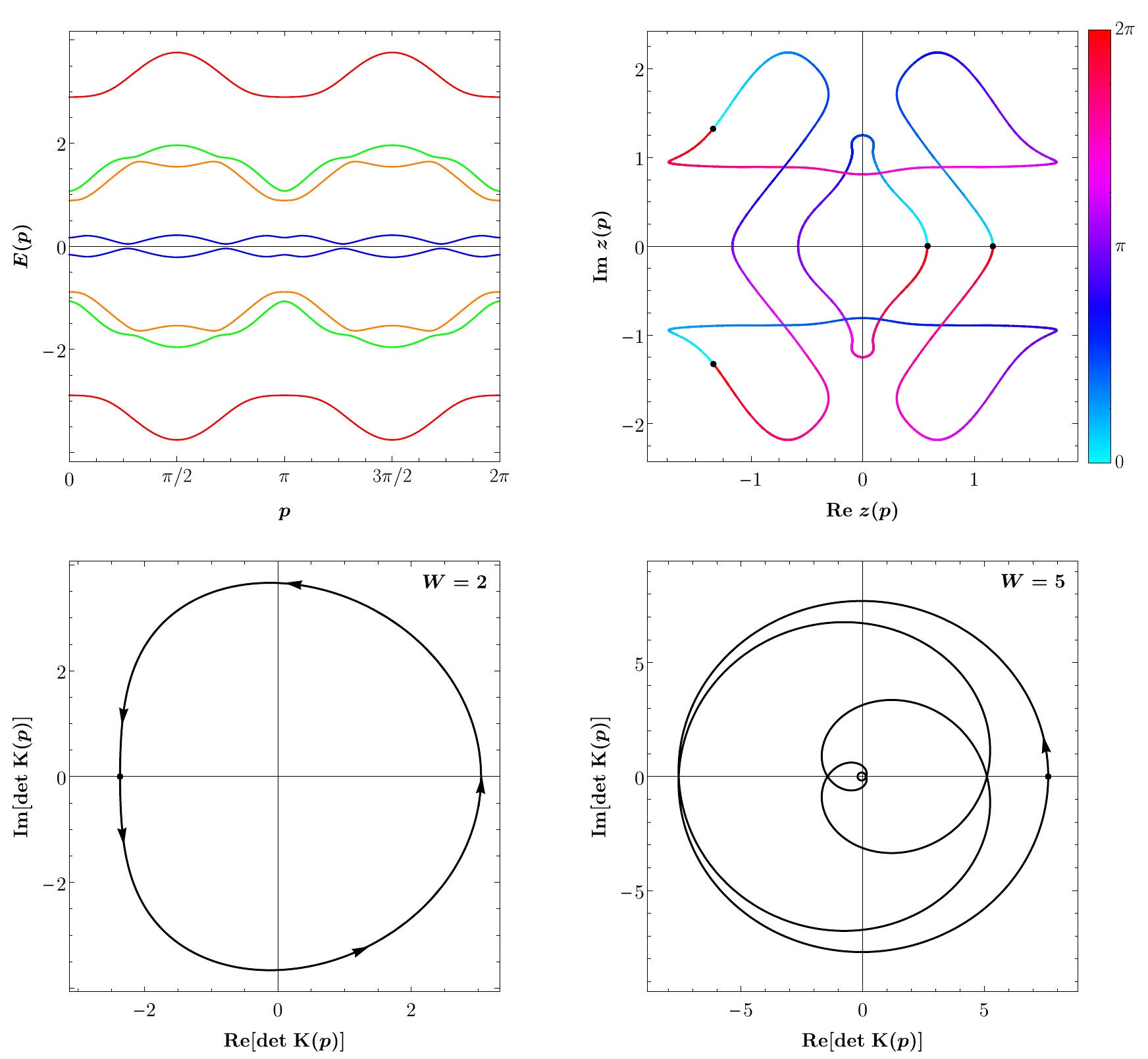}
\caption{A realization of a BDI Hamiltonian with $K(p)=\cos(p)K_1+i\sin(p)K_2$ and some fixed $4\times4$ real matrices $K_1$ and $K_2$. The top left plot shows the real eigenvalues of $H(p)$, the top right one shows the generically complex eigenvalues of $K(p)$, and the bottom left plot depicts the determinant $\det K(p)$. The bottom right plot shows the determinant of a different random matrix field $K(p) = (a_1 + a_2 e^{ip} + a_3 e^{2ip}+ a_4 e^{3ip}) K_1 + (b_1 + b_2 e^{ip} + b_3 e^{2ip}+ b_4 e^{3ip}) K_2$ with fixed $4\times4$ real matrices $K_1$ and $K_2$ and real coefficients $a_j$ and $b_j$. All plots show the parametric dependence in $p\in[0,2\pi)$ where we have employed the step size $2\pi/100$ and a B-Spline to obtain the curves. In the parametric plots the starting points $p=0$ are marked by black points and the directions are marked by a color gradient resp. arrows.}
\label{2fig}
\end{figure}

The pertinent topological invariant in the chiral classes AIII, BDI and CII is the winding number of $\det\, K(p)$ around the origin. It can be expressed as the integral\cite{Maffei2018, AsbothBook}
\begin{equation} \label{2WindingNumberDef}
W = \frac{1}{2\pi i} \oint\limits_{\det K(p)} \frac{dz}{z} = \frac{1}{2\pi i} \int\limits_0^{2\pi} dp\, w(p),
\end{equation}
where the logarithmic derivative
\begin{equation} \label{2WindingNumberDensityDef}
w(p) = \frac{d}{dp} \ln \det K(p) = \frac{1}{\det K(p)} \frac{d}{dp} \det K(p)
\end{equation}
is the winding number density. The integral \eqref{2WindingNumberDef} is an integer as can be shown by invoking Cauchy's argument principle when writing the integral as a contour integral in $e^{ip}$, see Ref.~\onlinecite{Guhr2023}. The real- and imaginary parts of $w(p)$ have different interpretations. While the real part describes the radial part of $\det K(p)$ and always integrates to zero over a closed contour, the imaginary part describes its angle. Therefore the latter would be sufficient to define the winding number.

In Fig.~\ref{2fig} we illustrate the spectral flow of $K(p) = \cos(p) K_1 + i \sin(p) K_2$ with generic real $K_1, K_2 \in \mathbb{R}^{4\times 4}$. We plotted the eight real eigenvalues of the Hamiltonian $H(p)$, showing a reflection symmetry at $E=0$ due to chiral symmetry, and the four complex eigenvalues of $K(p)$ as well as the determinant $\det\, K(p)$. Due to level repulsion no degeneracies occur in the spectrum of $H(p)$, which implies that each eigenvalue is a $2\pi$-periodic function. This is not the case for the complex eigenvalues of $K(p)$, which may experience a permutation when running from $p=0$ to $p=2\pi$, as can be observed in Fig.~\ref{2fig}. However, the determinant of $K(p)$ is always a $2\pi$-periodic function and therefore suitable to define a winding number. For our specific choice of the parametric dependence we have $K(p+\pi) = -K(p)$, which restricts the amount of times $\det\, K(p)$ winds around the origin to be an even resp. odd number for even resp. odd matrix dimensions. This symmetry does not exist in general as we also illustrated in Fig.~\ref{2fig} by means of another random matrix field $K(p) = (a_1 + a_2 e^{ip} + a_3 e^{2ip}+ a_4 e^{3ip}) K_1 + (b_1 + b_2 e^{ip} + b_3 e^{2ip}+ b_4 e^{3ip}) K_2$ with generic real $K_1, K_2 \in \mathbb{R}^{4\times 4}$ and real coefficients $a_j$ and $b_j$. In this example we find an odd winding number.

The constraint \eqref{2parametricTRI} is also reflected in the spectral flow of the Hamiltonian when considering the symmetry class CII, like we did in Ref.~\onlinecite{HKGG2023}. The Bloch Hamiltonian becomes self-dual at the time reversal invariant momenta and therefore its eigenvalues show Kramers' degeneracy at these points on the parameter manifold.

The simplest choice for a non-trivial parametric dependence on a one-dimensional manifold is the random matrix field
\begin{equation} \label{2RMF}
K(p) = a(p) K_1 +b(p) K_2,
\end{equation}
where $a(p)$ and $b(p)$ are smooth $2\pi$-periodic and not simultaneously vanishing complex-valued functions and $K_1$ and $K_2$ are real Ginibre matrices. For notational reasons we gather the coefficient functions in a complex two-dimensional vector
\begin{equation}	\label{2vDef}
v(p)=(a(p),b(p)) \in\mathbb{C}^2\setminus \{0\}.
\end{equation}
In our case, where time reversal invariance is present, condition \eqref{2parametricTRI} translates to
\begin{equation}
v^*(p) = v(-p).
\end{equation}
However, the expressions we derive below remain valid for arbitrary non-zero $a(p)$ and $b(p)$. Therefore, our results may be of interest even if $p$ is not interpreted as a (quasi-)momentum.

We denote an average over the Ginibre matrices $K_1$ and $K_2$ as
\begin{equation} \label{2EnsAvg}
\langle F \rangle = \int d[K_1,K_2] P(K_1)P(K_2)F(K_1,K_2).
\end{equation}
Our goal is to compute the ensemble average over a ratio of determinants
\begin{equation} \label{2GenFct}
Z^{(1,N)}_{k|l}(q,p) = \left\langle \frac{\prod_{j=1}^l \det K(p_j)}{\prod_{j=1}^k \det K(q_j)} \right\rangle.
\end{equation}
The superscript refers to the Dyson index $\beta = 1$, i.e. it defines the symmetry class, and the dimension $N$ of the block matrix $K(p)$. Although we are only interested in the case $k=l$ we employ the more general definition with $k \neq l$ as this function can be traced back to averages of only two determinants for which $k+l=2$. This is shown in Sec.~\ref{secIV}, see also Refs.~\onlinecite{KieburgGuhr2010a,KieburgGuhr2010b} for more details and the derivation of this result. This quantity is the key to the winding number statistics in form of the $k$-point correlation of winding number densities
\begin{equation} \label{2kPointCorrDef}
C^{(1,N)}_k (p_1,\ldots,p_k) = \langle w(p_1) \cdots w(p_k) \rangle,
\end{equation}
generated by taking the derivative of \eqref{2GenFct}
\begin{equation} \label{2kPointCorrGen}
C^{(1,N)}_k(p_1,\ldots,p_k) = \frac{\partial^k}{\prod_{j=1}^k \partial p_j} Z^{(1,N)}_{k|k}(q,p)\Bigg|_{q=p}.
\end{equation}
We relegate the evaluation of these derivatives and taking the large $N$ limit in order to reveal universal properties of the statistics to a subsequent paper.

\section{Result}
\label{secIII}

It is well known that the orthogonal case (as well as the symplectic one) is described by a Pfaffian point process \cite{MehtaBook}. This means that the eigenvalue correlations can be written as a Pfaffian of an ensemble dependent kernel function. This will also be reflected in our results. Via the methods developed in Refs. \onlinecite{KieburgGuhr2010a, KieburgGuhr2010b} the ensemble average \eqref{2GenFct} for the two matrix model \eqref{2RMF} evaluates to a Pfaffian
\begin{equation}	\label{3PfRes}
Z^{(1,N)}_{k|k}(q,p) = \frac{1}{\det \left[\displaystyle \frac{1}{iv^T(q_m)\tau_2v(p_n)} \right]_{1\leq m,n\leq k}}
\Pf \begin{bmatrix}
\displaystyle \widehat{\rm K}_1(p_m,p_n) & \displaystyle \widehat{\rm K}_2(p_m,q_n)
\\
\displaystyle \overset{}{-\widehat{\rm K}_2(p_n,q_m)} & \displaystyle \widehat{\rm K}_3(q_m,q_n)
\end{bmatrix}_{1\leq m,n\leq k}
\end{equation}
of a $k\times k$-matrix with $2\times 2$ blocks. We chose the sign of the Pfaffian as
\begin{equation}
\Pf [i\tau_2,i\tau_2,\ldots,i\tau_2]=1,
\end{equation}
where $\tau_2$ is the second Pauli matrix. The blocks are the three kernel functions
\begin{equation}	\label{3KernRes}
\begin{split}
\widehat{\rm K}_1(p_m,p_n) =& \frac{N (N-1)}{4 \pi} \, iv^T(p_n) \tau_2 v(p_m)  \left[ v^T(p_m) v(p_n) \right]^{N-2},
\\
\widehat{\rm K}_2(p_n,q_m) =& \frac{1}{iv^T(p_n) \tau_2 v(q_m)} \frac{N (N-1)}{4\pi} \left(\frac{v^T(p_n) v(p_n)}{iv^T(p_n) \tau_2 v(q_m)}\right)^N 
\\
&\times \int\limits_{\mathbb{C}^2} d[z]\, z_1^{N-2} g^{(1,N)}(z_1,z_2)
\left(z_2 + \frac{v^T(p_n) v(q_m)}{iv^T(p_n) \tau_2 v(q_m)} \right)^{-1},
\\
\widehat{\rm K}_3(q_m,q_n) =& \frac{1}{[b(q_m) b(q_n)]^N} \Bigg[ \int\limits_{\mathbb{C}^2} d[z] \frac{g^{(1,N)}(z_1,z_2)}{(a(q_m) + b(q_m)\, z_1)(a(q_n) + b(q_n)\, z_2)}
\\
&- \frac{N (N-1)}{4\pi} \int\limits_{\mathbb{C}^{4}} d[z] \frac{(z_3 - z_1) (z_1 z_3 + 1)^{N-2}}{(a(q_m) + b(q_m)\, z_2)(a(q_n) + b(q_n)\, z_4)} g^{(1,N)}(z_1,z_2) g^{(1,N)}(z_3,z_4) \Bigg]
\end{split}
\end{equation}
which are calculated in Sec.~\ref{secIV}. We made use of the notation \eqref{2vDef}, which, except for the third kernel function, nicely reflects a symmetry of our random matrix field, that we will illuminate more closely in a moment. The antisymmetric function
\begin{equation}	\label{3gDef}
g^{(1,N)}(z_1, z_2) = \frac{\abs{z_2-z_1}}{z_2-z_1} \frac{B(1/2, (N+1)/2) \delta(y_1) \delta(y_2) +2 \delta(x_1-x_2) \delta(y_1+y_2) Q(z_1,z_1^*)}{\left[(1+z_1^2)(1+z_2^2)\right]^{(N+1)/2}}
\end{equation} 
directly emerges from our random matrix problem as part of a joint eigenvalue probability density. For the complex integration variables we use the notation $z_j = x_j +iy_j$ with $x_j, y_j \in \mathbb{R}$. Furthermore, a lower incomplete Beta function appears in our results
\begin{equation}
\begin{split}
Q(z,z^*) &= B\left( \frac{4y^2}{\abs{1+z^2}^2 + 4y^2};1/2,(N+1)/2 \right),
\\
B(x;a,b) &= \int\limits_x^1 dt\, t^{a-1} (1-t)^{b-1} = 2\int\limits_{\sqrt{\frac{x}{1-x}}}^\infty dt \frac{t^{2a-1}}{(1+t^2)^{a+b}}, \text{ with } x\in [0,1],	
\\
B(0;a,b) &= B(a,b) = \frac{\Gamma(a) \Gamma(b)}{\Gamma(a+b)}.
\end{split}
\end{equation}
Inserting \eqref{3gDef} into the second and third kernels yields the full expressions
\begin{equation}
\begin{split}
\widehat{\rm K}_2(p_n,q_m) =& \frac{1}{iv^T(p_n) \tau_2 v(q_m)} \frac{N (N-1)}{2\pi} \left(\frac{v^T(p_n) v(p_n)}{iv^T(p_n) \tau_2 v(q_m)}\right)^N
\\
&\hspace*{-1cm}\times \Bigg[\frac{(-1)^{N/2} \pi\, B\left(1/2,\frac{N+1}{2}\right)}{N-1} \binom{(N-1)/2}{N} { _2F_1}\left(1,\frac{N+1}{2};N+1;1+\left( \frac{v^T(p_n) v(q_m)}{iv^T(p_n) \tau_2 v(q_m)} \right)^2\right)
\\
&\hspace*{-1cm}+ i \int\limits_\mathbb{C} d[z] \frac{z^{N-2}\, \sgn(\Im\, z) Q(z,z^*)}{\abs{1+z^2}^{N+1}} \left( z^* + \frac{v^T(p_n) v(q_m)}{iv^T(p_n) \tau_2 v(q_m)} \right)^{-1} \Bigg]
\end{split}
\end{equation}
with the hypergeometric function ${ _2F_1}$ and
\begin{equation}
\begin{split}
\widehat{\rm K}_3(q_m,q_n) =& \frac{1}{[b(q_m) b(q_n)]^N} \left[\int\limits_\mathbb{R} dx \frac{r(x,v(q_n))}{(a(q_m)+b(q_m)x)} + \int\limits_\mathbb{C} d[z] \frac{s(z,z^*,v(q_n))}{(a(q_m)+b(q_m)z)}\right]
\\
&- \frac{N(N-1)}{2\pi [b(q_m) b(q_n)]^{N-1}} 
\Bigg[\int\limits_{\mathbb{R}^2} d[x] r(x_1,v(q_m)) r(x_2,v(q_n)) (x_2-x_1) (x_1x_2+1)^{N-2}
\\
&+\int\limits_\mathbb{R} dx \int\limits_\mathbb{C} d[z] \det\begin{bmatrix}
r(x,v(q_m)) & s(z,z^*,v(q_m))
\\
r(x,v(q_n)) & s(z,z^*,v(q_n))
\end{bmatrix}
(z-x) (zx+1)^{N-2}
\\
&+\int\limits_{\mathbb{C}^2} d[z] s(z_1,z_1^*,v(q_m)) s(z_2,z_2^*,v(q_n)) (z_2-z_1) (z_1z_2+1)^{N-2} \Bigg]
\end{split}
\end{equation}
with the functions
\begin{equation}	\label{3rsfunctions}
\begin{split}
r(x,v(q)) &= B(1/2,(N+1)/2) \int\limits_{\mathbb{R}} dx' \frac{\sgn(x'-x)}{(a(q)+b(q)x') [(1+x^2) (1+x'^2)]^{(N+1)/2}},
\\
s(z,z^*,v(q)) &= \frac{2i\, \sgn(\Im\, z) Q(z,z^*)}{(a(q) + b(q) z^*) \abs{1+z^2}^{N+1}}.
\end{split}
\end{equation}
These expressions still contain integrals over up to four real dimensions. A further analytic computation of them is difficult due to the singularities appearing at various points in the complex plane. However, an evaluation in the large $N$ limit, which is our ultimate goal for the correlation function \eqref{2kPointCorrDef}, should be feasible.

\section{Derivation}
\label{secIV}

We start by highlighting a symmetry of the parametric matrix model \eqref{2RMF}, that we will use to simplify the necessary integrals. We continue by tracing the averages over the Ginibre ensembles back to an average over another random matrix ensemble. Then we give the expressions for the three kernel functions, as obtained from the supersymmetry without supersymmetry technique and calculate those one by one.

The matrices $K_1$ and $K_2$ are drawn from a real Ginibre distribution, i.e. their joint matrix distribution is
\begin{equation}
P(K_1) P(K_2) = (2\pi)^{-N^2} \exp\left[-\frac{1}{2}\tr\left(K_1 K_1^T+K_2 K_2^T\right)\right].
\end{equation}
This function is invariant under O(2) transformations
\begin{equation}	\label{4Sym}
\widehat{K} = \begin{bmatrix}
K_1 \\ K_2
\end{bmatrix} \to \left[ U \otimes \mathds{1}_N \right] 
\begin{bmatrix}
K_1 \\ K_2
\end{bmatrix},
\end{equation}
where $U \in \text{O}(2)$ acts on the two component matrix valued vector $\widehat{K}$. This invariance is inherited by the vectors $v(p)$
\begin{equation}
K(p) = a(p) K_1 +b(p) K_2 = v^T(p) \widehat{K}.
\end{equation}
Assuming for the moment $b(p) \neq 0$, we may rephrase the random matrix field
\begin{equation} \label{4KRephrasing}
K(p)=a(p)K_1 + b(p)K_2 = b(p)K_1\left(\kappa(p)\mathds{1}_N+K_1^{-1}K_2\right),\ {\rm with}\ \kappa(p)=\frac{a(p)}{b(p)},
\end{equation}
such that the generating function becomes
\begin{equation}	\label{4GenFunc}
Z^{(1,N)}_{k|k}(q,p) =  \left( \prod_{j=1}^k \frac{b(p_j)}{b(q_j)} \right)^N \left\langle \prod_{j=1}^k \frac{\det(\kappa(p_j)\mathds{1}_N+K_1^{-1}K_2)}{\det(\kappa(q_j)\mathds{1}_N+K_1^{-1}K_2)} \right\rangle.
\end{equation}
The random matrix $Y=K_1^{-1}K_2$ defines the real spherical ensemble, which was studied in Ref.~\onlinecite{ForresterMays2012}. The spherical ensemble is also well studied for the complex and the quaternion case, see Refs.~\onlinecite{Krishnapur2009, Mays2013}, which contain the results we used for our studies in the cases AIII and CII\cite{BHWGG2021, HKGG2023}. Its matrix probability density function in the real case is
\begin{equation}	\label{4mpdf}
\widetilde{G}^{(1,N)}(Y) = \pi^{-N^2/2} \prod_{j=1}^N \frac{\Gamma ((N+j)/2)}{ \Gamma(j/2)} \frac{1}{\det^N\left( \mathds{1}_N + YY^T\right)}.
\end{equation}
Unfortunately, in the literature the joint probability density of the eigenvalues $(z_1, \ldots, z_N) \in [\mathbb{C}\setminus \{0\}]^N$ exists only in a form, that is impractical for our purpose. Therefore we derive in Appendix \ref{A1} the following expression for $N$ even
\begin{equation}	\label{4jpdf}
\begin{split}
G^{(1,N)}(z) &= C^{(N)} \Delta_N(z) \prod_{j=1}^{N/2} g^{(1,N)}(z_{2j-1}, z_{2j}),
\\
C^{(N)} &= \frac{2^{-N/2} \pi^{-N/4}}{(N/2)!}
\prod_{j=1}^N \frac{\Gamma ((N+j)/2)}{\Gamma(j/2)}
\prod_{j=1}^{N/2} \frac{\Gamma\left( (N+1)/2 \right) \Gamma\left( N/2+1 \right)}{\Gamma\left( N+1/2-j \right) \Gamma\left( N+1-j \right)}
\end{split}
\end{equation}
with the antisymmetric function \eqref{3gDef}. The eigenvalues of a real matrix are either real or come in complex conjugated pairs, which is what \eqref{3gDef} describes. Although the part of this function concerning the complex eigenvalues is complex valued, the full joint eigenvalue probability density will be real as the corresponding phases drop out when considering the product with the Vandermonde determinant.

Only in the odd dimensional case we have to supplement this with an additional function for the unpaired eigenvalue, that must be real. This would complicate our already cumbersome expressions even more, which is why we decided to limit ourselves to the even case. This restriction is not an obstacle to our plan to deduce universal statistics in the large $N$ limit as the number of expected real eigenvalues is asymptotically given by $\sqrt{\pi N/2}$, regardless of the parity of $N$ \cite{ForresterMays2012}.

In the supersymmetry without supersymmetry technique \cite{KieburgGuhr2010a, KieburgGuhr2010b} one starts from integrals in a joint eigenvalue density. These can be reformulated by introducing superspace Jacobians, which in the present case are mixtures of Vandermonde and Cauchy determinants. This allows the application of a generalized version of de Bruijn's integration theorem \cite{deBruijn1955, KieburgGuhr2010b} resulting in a high dimensional block structured Pfaffian. After reducing the dimensionality by working out the Schur complement, the blocks of the Pfaffian can be identified as simplified averages over only two determinants.

Applying this method to \eqref{4GenFunc} yields
\begin{equation}
Z^{(1,N)}_{k|k}(q,p) = \frac{1}{\sqrt{\Ber_{k|k}^{(2)}(\kappa)}} 
\Pf \begin{bmatrix}
\displaystyle {\rm K}_1^{(1)}(p_m,p_n) & \displaystyle {\rm K}_2^{(1)}(p_m,q_n)
\\
\displaystyle \overset{}{-{\rm K}_2^{(1)}(p_n,q_m)} & \displaystyle \overset{}{{\rm K}_3^{(1)}(q_m,q_n)}
\end{bmatrix}_{\leq m,n\leq k},
\end{equation}
where the kernel functions are the simplified averages
\begin{equation}	\label{4Kernels}
\begin{split}
{\rm K}_1^{(1)}(p_m,p_n) &= [\kappa(p_n)-\kappa(p_m)] [b(p_m)b(p_n)]^N \widetilde{Z}_{0|2}^{(1,N-2)}(p_m,p_n),
\\
{\rm K}_2^{(1)}(p_n,q_m) &= \left(\frac{b(p_n)}{b(q_m)}\right)^N \frac{\widetilde{Z}_{1|1}^{(1,N)}(q_m,p_n)}{\kappa(q_m)-\kappa(p_n)} = \frac{Z_{1|1}^{(1,N)}(q_m,p_n)}{\kappa(q_m)-\kappa(p_n)},
\\
{\rm K}_3^{(1)}(q_m,q_n) &= \frac{\kappa(q_n)-\kappa(q_m)}{[b(q_m)b(q_n)]^N} \widetilde{Z}_{2|0}^{(1,N+2)}(q_m,q_n).
\end{split}
\end{equation}
We define for $l-k$ even and $2M+l-k \leq N$
\begin{equation}	\label{4ZtildeDef}
\begin{split}
\widetilde{Z}^{(1,2M)}_{k|l}(q,p) =& \frac{1}{M!\, \Pf\, D^{(2M+l-k)}} 
\\
&\times \int\limits_{\mathbb{C}^{2M}} d[z] \Delta_{2M}(z) \prod_{j=1}^{M} g^{(1,N)}(z_{2j-1},z_{2j}) \prod_{j=1}^{2M} \frac{\prod_{n=1}^l (\kappa(p_n)+z_j)}{\prod_{m=1}^k (\kappa(q_m)+z_j)}
\end{split}
\end{equation}
with the moment matrix
\begin{equation}	\label{4MomentMatDef}
\begin{split}
D^{(d)} &= \left[ D_{ab} \right]_{1\leq a,b \leq d},
\\
D_{ab} &= \int d[z] g^{(1,N)}(z_1,z_2)\, (z_1^{a-1} z_2^{b-1} - z_1^{b-1} z_2^{a-1}) = 2\int d[z] g^{(1,N)}(z_1,z_2)\, z_1^{a-1} z_2^{b-1}.
\end{split}
\end{equation}
The Berezinian is a superspace analogue to the Jacobian. In this case it is given by a Cauchy determinant\cite{Guhr1991, KieburgGuhr2010a}
\begin{equation}
\sqrt{\Ber_{k|k}^{(2)}(\kappa)} = \det\left[ \displaystyle\frac{1}{\kappa(q_m) - \kappa(p_n)} \right]_{1\leq a,b\leq k}.
\end{equation}
In the following subsections we calculate more explicit expressions of the kernels \eqref{4Kernels}. Furthermore, to arrive at the results \eqref{3PfRes} and \eqref{3KernRes} one has to rearrange some factors in order to highlight the symmetry \eqref{4Sym} of our problem.

\subsection{The Kernel \texorpdfstring{${\rm K}_1^{(1)}$}{}}

The first kernel is determined by $\widetilde{Z}_{0|2}^{(1,N-2)}(p_m,p_n)$, which is related to $Z_{0|2}^{(1,N-2)}(p_m,p_n)$. For the latter function we can use the O(2)-symmetry to reduce the amount of determinants in the average by one. We obtain the proper normalization via the limits
\begin{equation}
\lim_{\kappa(p) \to \infty} \frac{\widetilde{Z}_{0|2}^{(1,N-2)}(p_m,p_n)}{[\kappa(p_m) \kappa(p_n)]^{N-2}} = \frac{\Pf\, D^{(N-2)}}{\Pf\, D^{(N)}},	\qquad	\qquad	\lim_{a(p) \to \infty} \frac{Z_{0|2}^{(1,N-2)}(p_m,p_n)}{[a(p_m) a(p_n)]^{N-2}} = \left\langle \det\nolimits^2 K_1\right\rangle.
\end{equation}
These limits are equivalent because $\kappa(p) = a(p)/b(p)$ is directly proportional to $a(p)$, see \eqref{4KRephrasing}. Therefore the functions are related by
\begin{equation}
\widetilde{Z}_{0|2}^{(1,N-2)}(p_m,p_n) = \frac{\Pf\, D^{(N-2)}}{\Pf\, D^{(N)}} \frac{1}{\left\langle \det^2 K_1\right\rangle} \frac{Z_{0|2}^{(1,N-2)}(p_m,p_n)}{\left[ b(p_m)b(p_n) \right]^{N-2}}.
\end{equation}
Here and also in the following equation, the angular brackets denote an average over the $(N-2)$-dimensional real Ginibre ensemble, see Eq.~\eqref{2EnsAvg}. Now let $a_1, b_1, a_2, b_2 \in \mathbb{R}$ and
\begin{equation}	\label{4Xi1}
\Xi_1 = \frac{\left\langle \det\left( a_1 K_1+b_1 K_2 \right) \det\left( a_2 K_1+b_2 K_2 \right)\right\rangle}{\left\langle \det^2 K_1 \right\rangle}.
\end{equation}
This function is a polynomial in $a_1, b_1, a_2, b_2$ and can thus be analytically continued to the partition function $Z_{0|2}^{(1,N-2)}(p_m,p_n)$, that has, in general, complex coefficients. We are now finally in the position to exploit the O(2)-symmetry as we are allowed to rotate only real vectors. We use
\begin{equation}	\label{4UMat}
U = \frac{1}{\sqrt{a_1^2 + b_1^2}} \begin{bmatrix}
a_1 & -b_1
\\
b_1 & a_1
\end{bmatrix} \in \text{SO}(2)
\end{equation}
yielding
\begin{equation}
\Xi_1 = \frac{(a_1b_2-b_1a_2)^{N-2}}{\left\langle \det^2 K_1\right\rangle} \int\limits d[K_1] d[K_2] P(K_1) P(K_2) \det\nolimits^2 K_1 \det\left(\frac{a_1 a_2 + b_1 b_2}{a_1b_2-b_1a_2} + K_1^{-1}K_2\right).
\end{equation}
Up to the factor $\det^2 K_1$ appearing in the integrand this is almost an integral over the spherical ensemble. In fact, this type of measure leads to a generalized form of the spherical ensemble of matrices $K_1^{-1} K_2$, where one or both of $K_1$ and $K_2$ are drawn from a deformed Ginibre ensemble
\begin{equation}
P_\mu(K) = 2^{-N\mu} (2\pi)^{-N^2/2} \prod_{j=1}^N \frac{\Gamma(j/2)}{\Gamma(\mu + j/2)} \exp \left(-\frac{1}{2}\tr\, K K^T\right) \det\nolimits^{\mu} K K^T.
\end{equation}
Matrix ensembles of such kind are called induced spherical ensembles. They are well studied for $\beta = 1,2,4$, see Refs.~\onlinecite{FischmannForrester2011, MaysPonsaing2017, Fischmann2015}. In appendix \ref{A2} we show that the characteristic polynomial of the real spherical ensemble, induced or not, is given by a monomial. Thus, we find
\begin{equation}
\Xi_1 = (a_1 a_2 + b_1 b_2)^{N-2}
\end{equation}
and the partition function $Z_{0|2}^{(1,N-2)}(p_m,p_n)$ has the form
\begin{equation}
\frac{Z_{0|2}^{(1,N-2)}(p_m,p_n)}{\left\langle \det^2 K_1 \right\rangle} = [a(p_m) a(p_n) + b(p_m) b(p_n)]^{N-2} = v^T(p_m) v(p_n),
\end{equation}
which is indeed O(2)-invariant. The Pfaffians of the moment matrix can be related to the normalization of an induced spherical ensemble. This is covered in appendix \ref{A1}. We find
\begin{equation}	\label{4PfaffQuotient}
\frac{\Pf\, D^{(N-2)}}{\Pf\, D^{(N)}} = \frac{N (N-1)}{8 \pi}
\end{equation}
and altogether for the first kernel
\begin{equation}
\begin{split}
{\rm K}_1^{(1)}(p_m,p_n) =& \frac{N (N-1)}{8 \pi} b(p_m) b(p_n) [ a(p_n)b(p_m) - a(p_m)b(p_n) ] 
\\
&\times [a(p_m) a(p_n) + b(p_m) b(p_n)]^{N-2}
\\
=& \frac{N (N-1)}{8 \pi} b(p_m) b(p_n)\, iv^T(p_n) \tau_2 v(p_m) [v^T(p_m) v(p_n)]^{N-2},
\end{split}
\end{equation}
which by itself is not O(2)-invariant, just as the other kernels \eqref{4Kernels} will not be. However, the final expression for $Z_{k|k}^{(1,N)}(q,p)$ will be, because the non-invariant prefactor will be compensated.

\subsection{The Kernel \texorpdfstring{${\rm K}_2^{(1)}$}{}}

The second kernel is essentially the generator $Z_{1|1}^{(1,N)}(q_m,p_n)$ of the one-point function. Once again we want to apply the O(2)-symmetry to eliminate one of the determinants, for this purpose we consider
\begin{equation}	\label{4Xi2}
\Xi_2 = \left\langle \frac{\det\left( a_1 K_1+b_1 K_2 \right)}{ \det\left( a_2 K_1+b_2 K_2 \right)}\right\rangle.
\end{equation}
This function is a polynomial in $a_1, b_1$ and non-holomorphic in $a_2, b_2$. Therefore we may apply our arguments only to the numerator and have to set $a_1, b_1 \in \mathbb{R}$ and $a_2, b_2 \in \mathbb{C}$. We rotate with the same matrix \eqref{4UMat} as for the first kernel and obtain an integral over the spherical ensemble
\begin{equation}
\Xi_2 = \left(\frac{a_1^2+b_1^2}{a_1b_2-b_1a_2}\right)^N \int\limits d[Y] \widetilde{G}^{(1,N)}(Y) \frac{1}{\det\left(  \widehat{\kappa} + Y\right)}
\end{equation}
with
\begin{equation}
\widehat{\kappa} = \frac{a_1a_2+b_1b_2}{a_1b_2-b_1a_2}.
\end{equation}
We want to perform the integral in the joint probability density of the eigenvalues
\begin{equation}
\Xi_2 = C^{(N)} \left(\frac{a_1^2+b_1^2}{a_1b_2-b_1a_2}\right)^N \int\limits_{\mathbb{C}^N} d[z] \Delta_N(z) \prod_{j=1}^{N/2} g^{(1,N)}(z_{2j-1},z_{2j}) \prod_{j=1}^N \frac{1}{\widehat{\kappa}+z_j}.
\end{equation}
First, we identify part of the integrand as a Berezinian
\begin{equation}	\label{4BerDetIdentity}
\Delta_N(z) \prod_{j=1}^N \frac{1}{\widehat{\kappa} + z_j} = \sqrt{\Ber^{(2)}_{N|1} \left( z;-\widehat{\kappa} \right)} =
(-1)^{N+1} \det\left[\begin{array}{c|c}
\displaystyle z_a^{b-1} & \displaystyle \frac{1}{z_a+\widehat{\kappa}}
\end{array}\right]_{\substack{1\leq a\leq N \\ 1\leq b\leq N-1}},
\end{equation}
which in turn can be written as a determinant \cite{KieburgGuhr2010a}. We expand this determinant in the last column. Using its skew-symmetry under row permutation and the antisymmetry of $g^{(1,N)}(z_{2j-1},z_{2j})$ we find that each terms yields the same contribution, which is
\begin{equation}	\label{4Xi2Reduced}
\begin{split}
\Xi_2 &= (-1)^{N+1} N\, C^{(N)} \left(\frac{a_1^2+b_1^2}{a_1b_2-b_1a_2}\right)^N \int\limits_{\mathbb{C}^N} d[z] \frac{\Delta_{N-1}(z_1,\ldots,z_{N-1})}{z_N+\widehat{\kappa}} \prod_{j=1}^{N/2} g^{(1,N)}(z_{2j-1},z_{2j})
\\
&= -\frac{N(N-1)}{4\pi} \left(\frac{a_1^2+b_1^2}{a_1b_2-b_1a_2}\right)^N \int\limits_{\mathbb{C}^2} d[z] \frac{z_1^{N-2}}{z_2+\widehat{\kappa}} g^{(1,N)}(z_1,z_2).
\end{split}
\end{equation}
The integral over $z_1, \ldots , z_{N-2}$ can be conceived as a characteristic polynomial of an induced spherical ensemble in the variable $z_{N-1}$, which gives up to a prefactor a monomial. The details of this calculation are covered in appendix \ref{A2}. The integral over the remaining two eigenvalues yields two contributions, one for the case in which they are real and one for the case in which they are a complex conjugate pair
\begin{equation}	\label{4Xi2Res}
\Xi_2 = -\frac{N (N-1)}{4\pi} \left(\frac{a_1^2+b_1^2}{a_1b_2-b_1a_2}\right)^N [I_R(\widehat{\kappa}) + I_C(\widehat{\kappa})].
\end{equation}
The contribution of the real eigenvalues is
\begin{equation}	\label{4RealContribution}
\begin{split}
I_R(\widehat{\kappa}) &= B(1/2,(N+1)/2) \int\limits_{\mathbb{R}^2} d[x] \frac{\sgn (x_2-x_1) }{\left[(1+x_1^2)(1+x_2^2)\right]^{(N+1)/2}} \frac{x_1^{N-2}}{x_2 +\widehat{\kappa}}
\\
&= B(1/2,(N+1)/2) \int\limits_{x_2>x_1} d[x] \frac{1}{\left[(1+x_1^2)(1+x_2^2)\right]^{(N+1)/2}} \left( \frac{x_1^{N-2}}{x_2+\widehat{\kappa}} - \frac{x_2^{N-2}}{x_1+\widehat{\kappa}} \right),
\end{split}
\end{equation}
where we treat the sign function by splitting the integral into two terms, that we calculate separately. In the first term we integrate $x_1$ over $(-\infty, x_2]$ yielding
\begin{equation}
\begin{split}
&\int\limits_{-\infty}^\infty dx_2 \int\limits_{-\infty}^{x_2} dx_1 \frac{1}{\left[(1+x_1^2)(1+x_2^2)\right]^{(N+1)/2}} \frac{x_1^{N-2}}{x_2+\widehat{\kappa}}
\\
= &\int\limits_{-\infty}^\infty dx_2 \frac{1}{(1+x_2^2)^{(N+1)/2}} \frac{1}{x_2+\widehat{\kappa}} \frac{(-1)^N+x_2^{N-1}(1+x_2^2)^{(1-N)/2}}{N-1}
\end{split}
\end{equation}
and in the second term we integrate $x_2$ over $[x_1,\infty)$
\begin{equation}
\begin{split}
&\int\limits_{-\infty}^\infty dx_1 \int\limits_{x_1}^\infty dx_2 \frac{1}{\left[(1+x_1^2)(1+x_2^2)\right]^{(N+1)/2}} \frac{x_2^{N-2}}{x_1+\widehat{\kappa}}
\\
= &\int\limits_{-\infty}^\infty dx_1 \frac{1}{(1+x_1^2)^{(N+1)/2}} \frac{1}{x_1+\widehat{\kappa}} \frac{1-x_1^{N-1}(1+x_1^2)^{(1-N)/2}}{N-1}.
\end{split}
\end{equation}
For both integrals the antiderivative is
\begin{equation}
\int dx \frac{x^{N-2}}{(1+x^2)^{(N+1)/2}} = \frac{1}{N-1} \frac{x^{N-1}}{(1+x^2)^{(N-1)/2}}.
\end{equation}
Considering $N$ even one of the terms drops out and \eqref{4RealContribution} results in an integral over the real axis
\begin{equation}
I_R(\widehat{\kappa}) = \frac{2\, B(1/2,(N+1)/2)}{N-1} \int\limits_{\mathbb{R}} dx \frac{x^{N-1}}{(1+x^2)^{N}} \frac{1}{x+\widehat{\kappa}}.
\end{equation}
In the case that $\widehat{\kappa}$ is real it has to be understood as a principal value integral. We continue by symmetrizing the integrand and substituting $t = x^2$
\begin{equation}
\begin{split}
I_R(\widehat{\kappa}) &= \frac{2\, B(1/2,(N+1)/2)}{N-1} \int\limits_0^\infty dx \frac{x^{N-1}}{(1+x^2)^{N}} \left(\frac{1}{x+\widehat{\kappa}} + \frac{1}{x-\widehat{\kappa}} \right)
\\
&= \frac{2\, B(1/2,(N+1)/2)}{N-1} \int\limits_0^\infty dt \frac{t^{(N-1)/2}}{(1+t)^{N}} \frac{1}{t-\widehat{\kappa}^2}.
\end{split}
\end{equation}
The last integrand has a branch cut along the negative real axis. It is equivalent to an integral of the function
\begin{equation}
f(z) = \frac{z^{(N-1)/2}}{(1-z)^N} \frac{-1}{z+\widehat{\kappa}^2}
\end{equation}
over a keyhole contour around the negative real axis. Applying the residue theorem yields a truncated binomial series
\begin{equation}
\begin{split}
I_R(\widehat{\kappa}) &= \frac{(-1)^{N/2} 2\pi\, B(1/2,(N+1)/2)}{N-1} \left[ \sum_{l=0}^{N-1} \binom{(N-1)/2}{l} \frac{(-1)^{N-1-l}}{(1+\widehat{\kappa}^2)^{N-l}} + \frac{(-\widehat{\kappa}^2)^{(N-1)/2}}{(1+\widehat{\kappa}^2)^N} \right]
\\
&= \frac{(-1)^{N/2} 2\pi\, B(1/2,(N+1)/2)}{N-1} \binom{(N-1)/2}{N} { _2F_1}\left(1,(N+1)/2;N+1;1+\widehat{\kappa}^2\right).
\end{split}
\end{equation}
The second representation involves the hypergeometric function \cite{NIST}. The contribution of the complex conjugated eigenvalues is
\begin{equation}	\label{4ComplexContribution}
I_C(\widehat{\kappa}) = 2i \int\limits_\mathbb{C} d[z] \sgn(\Im\, z) \frac{z^{N-2}}{z^*+\widehat{\kappa}} \frac{Q(z,z^*)}{\abs{1+z^2}^{N+1}}.
\end{equation}
Therefore, we find for the second kernel
\begin{equation}
\begin{split}
{\rm K}_2^{(1)}(p_n,q_m) =& \frac{N (N-1)}{2\pi} \frac{b(p_n) b(q_m)}{iv^T(p_n) \tau_2 v(q_m)} \left( \frac{v^T(p_n) v(p_n)}{iv^T(p_n) \tau_2 v(q_m)} \right)^N
\\
&\hspace*{-1cm}\times \Bigg[ \frac{(-1)^{N/2} \pi B\left(1/2,\frac{N+1}{2}\right)}{N-1} \binom{(N-1)/2}{N} { _2F_1}\left(1,\frac{N+1}{2};N+1;1+\left( \frac{v^T(p_n) v(q_m)}{iv^T(p_n)\tau_2 v(q_m)} \right)^2\right)
\\
&\hspace*{-1cm}+ i \int\limits_\mathbb{C} d[z] \sgn(\Im\, z) \frac{z^{N-2} Q(z,z^*)}{\abs{1+z^2}^{N+1}} \left( z^* + \frac{v^T(p_n) v(q_m)}{iv^T(p_n)\tau_2 v(q_m)} \right)^{-1} \Bigg].
\end{split}
\end{equation}
Once again we note the O(2)-invariance of this function in the vectors $v(p) = (a(p),b(p))$ up to a prefactor.

\subsection{The Kernel \texorpdfstring{${\rm K}_3^{(1)}$}{}}

The third kernel is determined by $\widetilde{Z}^{(1,N+2)}_{2|0}(q_m,q_n)$. Therefore we need to evaluate the following integral
\begin{equation}
\Xi_3 = \frac{1}{(N/2+1)!\, \Pf\, D^{(N)}} \int\limits_{\mathbb{C}^{N+2}} d[z] \Delta_{N+2}(z) \prod_{j=1}^{N/2+1} g^{(1,N)}(z_{2j-1},z_{2j}) \prod_{j=1}^{N+2} \frac{1}{(\kappa_1+z_j)(\kappa_2+z_j)}.
\end{equation}
This time we cannot use the O(2)-symmetry to simplify the integral, because it cannot be traced back to the ensemble average \eqref{2GenFct}. Instead, we proceed by applying the identity \cite{KieburgGuhr2010a}
\begin{equation}
\begin{split}
\Delta_{N+2}(z) \prod_{j=1}^{N+2} \frac{1}{(\kappa_1+z_j)(\kappa_2+z_j)} &= \frac{1}{\kappa_1-\kappa_2} \sqrt{\Ber^{(2)}_{N+2|2} \left( z;-\kappa \right)} 
\\
&= \frac{1}{\kappa_2-\kappa_1}
\det\left[\begin{array}{c|c|c}
\displaystyle z_a^{b-1} & \displaystyle \frac{1}{z_a+\kappa_1} & \displaystyle \frac{1}{z_a+\kappa_2}
\end{array}\right]_{\substack{1\leq a\leq N+2 \\ 1\leq b\leq N}}
\end{split}
\end{equation}
and expand the determinant in the last two columns. Similar to \eqref{4Xi2Reduced} we use the antisymmetry of $g^{(1,N)}(z_{2j-1},z_{2j})$ to reduce the amount of terms appearing in the integral
\begin{equation}	\label{4Xi3Reduced}
\begin{split}
\Xi_3 =& \frac{2\, C^{(N)}}{(N+2) (\kappa_2-\kappa_1)} \int\limits_{\mathbb{C}^{N+2}} d[z] \prod_{j=1}^{N/2+1} g^{(1,N)}(z_{2j-1},z_{2j}) 
\det\left[\begin{array}{c|c|c}
\displaystyle z_a^{b-1} & \displaystyle \frac{1}{z_a+\kappa_1} & \displaystyle \frac{1}{z_a+\kappa_2}
\end{array}\right]_{\substack{1\leq a\leq N+2 \\ 1\leq b\leq N}}
\\
=& \frac{2\, C^{(N)}}{\kappa_2-\kappa_1} \Bigg[ \int\limits_{\mathbb{C}^{N+2}} d[z] \frac{\Delta_N(z)}{(z_{N+1} + \kappa_1)(z_{N+2} + \kappa_2)} \prod_{j=1}^{N/2+1} g^{(1,N)}(z_{2j-1},z_{2j})
\\
&- N \int\limits_{\mathbb{C}^{N+2}} d[z] \frac{\Delta_N(z_1,\ldots, z_{N-1}, z_{N+1})}{(z_N + \kappa_1)(z_{N+2} + \kappa_2)} \prod_{j=1}^{N/2+1} g^{(1,N)}(z_{2j-1},z_{2j})
\Bigg].
\end{split}
\end{equation}
In the first term the pair of eigenvalues $z_{N+1}, z_{N+2}$ is decoupled from $z_1, \ldots, z_N$. Integration over the latter yields only a constant
\begin{equation}
C^{(N)} \int\limits_{\mathbb{C}^{N+2}} d[z] \frac{\Delta_N(z)}{(z_{N+1} + \kappa_1)(z_{N+2} + \kappa_2)} \prod_{j=1}^{N/2+1} g^{(1,N)}(z_{2j-1},z_{2j}) = \int\limits_{\mathbb{C}^2} d[z] \frac{g^{(1,N)}(z_1,z_2)}{(z_1 + \kappa_1)(z_2 + \kappa_2)}.
\end{equation}
In the second term we expand the Vandermonde determinant in the last two variables
\begin{equation}
\begin{split}
&\phantom{= } C^{(N)} \int\limits_{\mathbb{C}^{N+2}} d[z] \frac{\Delta_N(z_1,\ldots, z_{N-1}, z_{N+1})}{(z_N + \kappa_1)(z_{N+2} + \kappa_2)} \prod_{j=1}^{N/2+1} g^{(1,N)}(z_{2j-1},z_{2j})
\\
&= C^{(N)} \int\limits_{\mathbb{C}^{N+2}} d[z] \Delta_{N-2}(z) \frac{z_{N+1} - z_{N-1}}{(z_N + \kappa_1)(z_{N+2} + \kappa_2)}
\prod_{j=1}^{N-2} (z_{N+1}-z_j) (z_{N-1}-z_j)
\prod_{j=1}^{N/2+1} g^{(1,N)}(z_{2j-1},z_{2j}).
\end{split}
\end{equation}
This allows us to identify the integral over $z_1, \ldots z_{N-2}$ as the function $\Xi_1$, see Eq.~\eqref{4Xi1}, that we calculated for the first kernel, resulting in
\begin{equation}
\begin{split}
&\phantom{= } \frac{N-1}{4\pi} \int\limits_{\mathbb{C}^{4}} d[z] \frac{z_3 - z_1}{(z_2 + \kappa_1)(z_4 + \kappa_2)} g^{(1,N)}(z_1,z_2) g^{(1,N)}(z_3,z_4) \frac{\left\langle \det\left( z_1 K_1 - K_2 \right) \det\left( z_3 K_1 - K_2 \right)\right\rangle}{\left\langle \det^2 K_1\right\rangle}
\\
&= \frac{N-1}{4\pi} \int\limits_{\mathbb{C}^{4}} d[z] \frac{z_3 - z_1}{(z_2 + \kappa_1)(z_4 + \kappa_2)} g^{(1,N)}(z_1,z_2) g^{(1,N)}(z_3,z_4)\, (z_1 z_3 + 1)^{N-2}.
\end{split}
\end{equation}
Altogether the third kernel function is given by
\begin{equation}
\begin{split}
{\rm K}_3^{(1)}(q_m,q_n) =& \frac{2}{[b(q_m) b(q_n)]^{N-1}} \Bigg[ \int\limits_{\mathbb{C}^2} d[z] \frac{g^{(1,N)}(z_1,z_2)}{(a(q_m) + b(q_m)z_1)(a(q_n) + b(q_n)z_2)}
\\
&- \frac{N (N-1)}{4\pi} \int\limits_{\mathbb{C}^{4}} d[z] \frac{(z_3 - z_1) (z_1 z_3 + 1)^{N-2}}{(a(q_m) + b(q_m)z_2)(a(q_n) + b(q_n)z_4)} g^{(1,N)}(z_1,z_2) g^{(1,N)}(z_3,z_4) \Bigg].
\end{split}
\end{equation}
Unlike for the first and the second kernel in this result the O(2)-symmetry is not visible. However, our end result has to entail this symmetry. We define the functions
\begin{equation}
\begin{split}
r(x,v(q)) &= B(1/2,(N+1)/2) \int\limits_{\mathbb{R}} dx' \frac{\sgn(x'-x)}{(a(q)+b(q)x')\, [(1+x^2) (1+x'^2)]^{(N+1)/2}},
\\
s(z,z^*,v(q)) &= \frac{2i\, \sgn(\Im\, z) Q(z,z^*)}{(a(q) + b(q) z^*) \abs{1+z^2}^{N+1}}
\end{split}
\end{equation}
Inserting \eqref{3gDef} yields the cumbersome expression
\begin{equation}
\begin{split}
{\rm K}_3^{(1)}(q_m,q_n) =& \frac{2}{[b(q_m) b(q_n)]^{N-1}} \left[\int\limits_\mathbb{R} dx \frac{r(x,v(q_n))}{(a(q_m)+b(q_m)x )} + \int\limits_\mathbb{C} d[z] \frac{s(z,z^*,v(q_n))}{(a(q_m)+b(q_m)z)}\right]
\\
&- \frac{N(N-1)}{2\pi [b(q_m) b(q_n)]^{N-1}} 
\Bigg[\int\limits_{\mathbb{R}^2} d[x] r(x_1,v(q_m)) r(x_2,v(q_n)) (x_2-x_1) (x_1x_2+1)^{N-2}
\\
&+\int\limits_\mathbb{R} dx \int\limits_\mathbb{C} d[z] \det\begin{bmatrix}
r(x,v(q_m)) & s(z,z^*,v(q_m))
\\
r(x,v(q_n)) & s(z,z^*,v(q_n))
\end{bmatrix} (z-x) (zx+1)^{N-2}
\\
&+\int\limits_{\mathbb{C}^2} d[z] s(z_1,z_1^*,v(q_m)) s(z_2,z_2^*,v(q_n)) (z_2-z_1) (z_1z_2+1)^{N-2} \Bigg],
\end{split}
\end{equation}
which contains six integrals over at most four real variables.

\section{Conclusions}
\label{secV}

The winding number is the pertinent topological invariant for chiral symmetric Bloch Hamiltonians over a one-dimensional parameter manifold. We studied its statistics in the chiral orthogonal class BDI by calculating the ensemble averages for ratios of parametric determinants. This is the key quantity for the winding number statistics as it generates the correlation functions of the winding number densities. We did this by first mapping our problem to a spectral one, namely finding the ratios of characteristic polynomials in the spherical ensemble. Then we applied the supersymmetry without supersymmetry technique to write the ensemble average as a Pfaffian of three kernel functions, that are simplified averages over only two determinants. We could simplify the kernel function even further by employing a symmetry of our random matrix model. This complements our recent studies on the chiral unitary and chiral symplectic class \cite{BHWGG2021, HKGG2023}.

For technical reasons we considered only the case of even matrix dimensions. This does not interfere with our goal to study the universal behaviour in the limit of large matrix dimensions as the even and odd case behave equally in this limit. Universality is reached in the limit of large matrix dimension when also considering the parametric dependence on proper scales. In Ref.~\onlinecite{BHWGG2021} we additionally addressed this question for the correlation functions of winding number densities in the chiral unitary class AIII. Obtaining the universal limits for the correlation functions involves taking the derivatives of the here computed ensemble averages. Although the resulting expressions for the ensemble averages are quite cumbersome, we are positive that in the universality limit the correlation functions are feasible. We want to address this issue further in a future work.

\begin{acknowledgments}
  We thank Boris Gutkin for fruitful discussions.  This work was
  funded by the German--Israeli Foundation within the project
  \textit{Statistical Topology of Complex Quantum Systems}, grant
  number GIF I-1499-303.7/2019 (N.H.,O.G. and T.G.). Furthermore, M.K. acknowledges support by the Australian
Research Council via Discovery Project grant DP210102887.
\end{acknowledgments}

\appendix

\section{The joint eigenvalue probability density of the real induced spherical ensemble}	\label{A1}

We consider the real induced spherical ensemble with the matrix probability density
\begin{equation}	\label{A1mpdf}
\widetilde{G}^{(1,N)}_{\mu,\nu}(Y) = \pi^{-N^2/2} \prod_{j=1}^N \frac{\Gamma(j/2) \Gamma (\mu + \nu +(N+j)/2)}{\Gamma (\nu + j/2) \Gamma(\mu + j/2)} \frac{\det^{2\nu} Y}{\det^{N+\mu+\nu}\left( \mathds{1}_N + YY^T\right)}.
\end{equation}
It is the ensemble of matrices $Y = K_1^{-1}K_2$, where $K_1$ and $K_2$ are distributed according to deformed Gaussians
\begin{equation}	\label{A1DeformedGaussian}
P_\mu(K) = 2^{-N\mu} (2\pi)^{-N^2/2} \prod_{j=1}^N \frac{\Gamma(j/2)}{\Gamma(\mu + j/2)} \exp \left(-\frac{1}{2}\tr\, K K^T\right) \det\nolimits^{\mu} K K^T
\end{equation}
for the parameters $\mu$ resp. $\nu$. In the body of the text we omitted these indices, implying that they are zero, e.g. $\widetilde{G}^{(1,N)}_{0,0}(Y) = \widetilde{G}^{(1,N)}(Y)$, which leads to the ordinary spherical ensemble. The induced spherical ensemble is well-studied for $\beta = 1,2,4$ \cite{FischmannForrester2011, MaysPonsaing2017, Fischmann2015}. Unfortunately, in the real case the known joint eigenvalue probability density is available only in a form, that is impractical for our purpose. For this reason we will reproduce this result once again. In doing so, we adhere to methods employed in the aforementioned works. 

Generally for real matrices, we have to distinguish between even ($\widetilde{N} = N/2$) and odd ($\widetilde{N} = (N-1)/2$) matrix dimensions. We apply a real Schur decomposition
\begin{equation}
Y = U (D+T) U^T,
\end{equation}
where $D$ is a diagonal matrix of $\widetilde{N}$ real $2\times 2$-blocks $D_j$. In the odd case these are complemented by a real $1\times 1$-block $D_{\widetilde{N}+1}$. The matrix $T$ is a real strict upper triangular matrix and $U$ is orthogonal, $U \in \text{O}(N)/\text{O}(2)^{\widetilde{N}}$ resp. $U \in \text{O}(N)/(\text{O}(2)^{\widetilde{N}} \times \text{O}(1))$ for even resp. odd $N$. Upon this transformation the measure changes according to \cite{IpsenKieburg2014, Edelman1997}
\begin{equation}
d[Y] = \Delta_N(z) \prod_{j=1}^{\widetilde{N}} \frac{1}{z_{-j}-z_{+j}} d[D] d[T] d\mu(U),
\end{equation}
where $z_{+j}$ and $z_{-j}$ are the eigenvalues of $D_j$ and thus also of $Y$. First, one integrates over the upper triangular matrix $T$. Only the denominator of the matrix density \eqref{A1mpdf} depends on $T$. The determinant is expanded and the free parameters are integrated out column wise. In the even case this yields \cite{ForresterMays2012, Fischmann2015}
\begin{equation}
\begin{split}
\int d[T] \frac{\det^{2\nu} Y}{\det^{N+\mu+\nu}\left( \mathds{1}_N + YY^T\right)} =& \prod_{j=1}^{\widetilde{N}} \pi^{N-2j} \frac{\Gamma\left( N/2+\mu+\nu +1/2 \right) \Gamma\left( N/2+\mu+\nu+1 \right)}{\Gamma\left( N+\mu+\nu+1/2-j \right) \Gamma\left( N+\mu+\nu+1-j \right)}
\\
&\times \prod_{j=1}^{\widetilde{N}} \frac{\det^{2\nu}D_j}{\det^{N/2+\mu+\nu+1}(\mathds{1}_N+D_jD_j^T)}
\end{split}
\end{equation}
and in the odd case
\begin{equation}
\begin{split}
\int d[T] \frac{\det^{2\nu} Y}{\det^{N+\mu+\nu}\left( \mathds{1}_N + YY^T\right)} =& \prod_{j=1}^{\widetilde{N}} \pi^{N-2j} \frac{\Gamma\left( N/2+\mu+\nu +1/2 \right) \Gamma\left( N/2+\mu+\nu+1 \right)}{\Gamma\left( N+\mu+\nu+1/2-j \right) \Gamma\left( N+\mu+\nu+1-j \right)}
\\
&\times h_{\mu,\nu}^{(N)}(z_{2\widetilde{N}+1}) \prod_{j=1}^{\widetilde{N}} \frac{\det^{2\nu}D_j}{\det^{N/2+\mu+\nu+1}(\mathds{1}_N+D_jD_j^T)}.
\end{split}
\end{equation}
Here, we obtain an additional factor
\begin{equation}
h_{\mu,\nu}^{(N)}(z) = \frac{z^{2\nu}}{\left(1+z^2\right)^{N/2+\mu+\nu+1/2}} \delta\left( y \right)
\end{equation}
which effectively renders the variable $z_{2\widetilde{N}+1}$ real. We use the notation $z = x + i y$ with $x, y \in \mathbb{R}$ for the generally complex eigenvalues. Next, we work on the $2\times 2$-blocks. Gathering all factors depending on the eigenvalues of $D_j$, except those in the Vandermonde determinant, we obtain
\begin{equation}	\label{A12x2BlockMeasure1}
A_j = \frac{1}{z_{-j}-z_{+j}} \frac{\det^{2\nu}D_j}{\det^{N/2+\mu+\nu+1}(\mathds{1}_N+D_jD_j^T)}.
\end{equation}
Following Ref.~\onlinecite{SommersWieczorek2008}, we start by diagonalizing the symmetric part
\begin{equation}
O_j^T D_jO_J = \begin{bmatrix}
\lambda_{1j} & \rho_j
\\
-\rho_j & \lambda_{2j}
\end{bmatrix}	\qquad	
O_j = \begin{bmatrix}
\cos \varphi_j	& \sin \varphi_j
\\
-\sin \varphi_j & \cos \varphi_j
\end{bmatrix} \in \text{SO}(2).
\end{equation}
The range of parameters is $\rho_j \in \mathbb{R}$, $\varphi_j \in [0,\pi)$ and $\lambda_{1j},\lambda_{2j} \in \mathbb{R}$ with $\lambda_{1j} \geq \lambda_{2j}$. The flat measure of the four real independent variables transforms as
\begin{equation}
d[D_j] = 2(\lambda_{1j}-\lambda_{2j}) d\varphi_j d\rho_j d\lambda_{1j} d\lambda_{2j}.
\end{equation}
Next, we want to change coordinates from the eigenvalues of the symmetric part $\lambda_{1j},\lambda_{2j}$ to the eigenvalues $z_{\pm j}$ of the full matrix. They are related via
\begin{equation}	\label{A12x2Eigenvalues}
z_{\pm j} = \frac{\lambda_{1j}+\lambda_{2j}}{2} \pm \sqrt{\left(\frac{\lambda_{1j}-\lambda_{2j}}{2}\right)^2 - \rho_j^2}
\qquad
\lambda_{1,2 j} = \frac{z_{+j} + z_{-j}}{2} \pm \sqrt{\left(\frac{z_{+j} - z_{-j}}{2}\right)^2 + \rho_j^2}.
\end{equation}
According to the ordering of $\lambda_1$ and $\lambda_2$, the eigenvalues $z_{\pm j}$ also have to be ordered
\begin{equation}	\label{A1EVOrdering}
\begin{split}
z_{\pm j} \in \mathbb{R} &: x_{+j} \geq x_{-j}	
\\
z_{\pm j} = z^*_{\mp j} &: y_{+j} = - y_{-j} > 0,
\end{split}
\end{equation}
where we distinguished between the cases of real and complex conjugated eigenvalues. However, in the following we will disregard this ordering, which can be compensated by an overall factor $1/2$. We obtain the Jacobian
\begin{equation}
\abs{\det \frac{\partial(z_{+j}, z_{-j})}{\partial(\lambda_{1j}, \lambda_{2j})}} = \frac{\lambda_{1j}-\lambda_{2j}}{\abs{z_{-j}-z_{+j}}}.
\end{equation}
The determinants in the new coordinates are
\begin{equation}
\det\left( \mathds{1}_2 + D_j D_j^T \right) = 4\rho_j^2 + (1 + z_{+j}^2)(1 + z_{-j}^2)	\qquad	\det\, D_j = z_{+j} z_{-j}.
\end{equation}
We also need to consider that in the case, where $z_{\pm j}$ are complex conjugates, the integration regime of $\rho_j$ is confined to
\begin{equation}
\rho_j^2 \geq -\left( \frac{z_{+j} - z_{-j}}{2} \right)^2 = y^2_{\pm j}	\Rightarrow	\abs{\rho_j} \geq \abs{y_{\pm j}}.
\end{equation}
This is due to the condition $\lambda_{1,2 j} \in \mathbb{R}$. Gathering everything and integrating over $\varphi_j$ and $\rho_j$ we obtain for \eqref{A12x2BlockMeasure1}
\begin{equation}	\label{A12x2BlockMeasure2}
\begin{split}
\phantom{=}& 2(\lambda_{1j}-\lambda_{2j}) d\lambda_{1j} d\lambda_{2j} \int\limits_\mathbb{R} d\rho_j \int\limits_0^\pi d\varphi_j\, A_j 
\\
&= \frac{\pi}{2} \frac{\left(z_{+j} z_{-j}\right)^{2\nu}}{\left[(1 + z_+^2)(1 + z_-^2)\right]^{N/2+\nu+\mu+1/2}} \frac{\abs{z_{-j}-z_{+j}}}{z_{-j}-z_{+j}}
\\
&\times \bigg[ B(1/2,N/2+\mu+\nu+1/2) \Theta(x_{+j}-x_{-j}) \delta(y_{-j}) \delta(y_{+j})
\\
&+2\, \Theta(y_{+j}) \delta(x_{+j}-x_{-j}) \delta(y_{+j}+y_{-j}) Q_{\mu,\nu}^{(N)}(z_{+j}, z_{-j}) \bigg] dx_{+j} dx_{-j} dy_{+j} dy_{-j} ,
\end{split}
\end{equation}
where the function
\begin{equation}
\begin{split}
Q_{\mu,\nu}^{(N)}(z_{+j}, z_{-j}) &= 2 \int\limits_{\frac{2\abs{y_{\pm j}}}{\abs{1+z_{\pm j}^2}}}^\infty \frac{d\rho_j}{\left(1+\rho_j^2\right)^{N/2+\mu+\nu+1}} 
\\
&= B\left( \frac{4y^2}{\abs{1+z^2}^2 + 4y^2};1/2,N/2+\mu+\nu+1/2 \right)
\end{split}
\end{equation}
emerges. Now only the integral over the Haar measure of the respective cosets remains to be computed. It yields the constants
\begin{equation}
\begin{split}
\text{Vol}\left( \text{O}(N)/\text{O}(2)^{\widetilde{N}} \right) &= \frac{\text{Vol}(\text{O}(N))}{(4\pi)^{N/2}} = \frac{1}{(4\pi)^{N/2}} \prod_{j=1}^N \frac{2\pi^{j/2}}{\Gamma(j/2)},
\\
\text{Vol}\left( \text{O}(N)/(\text{O}(2)^{\widetilde{N}}\times \text{O}(1)) \right) &= \frac{\text{Vol}(\text{O}(N))}{2 (4\pi)^{(N-1)/2}} = \frac{1}{2 (4\pi)^{(N-1)/2}} \prod_{j=1}^N \frac{2\pi^{j/2}}{\Gamma(j/2)}
\end{split}
\end{equation}
for $N$ even resp. odd. We summarize this under
\begin{equation}
\int d\mu(U) = \frac{1}{(1+N-2\widetilde{N}) (4\pi)^{\widetilde{N}}} \prod_{j=1}^N \frac{2\pi^{j/2}}{\Gamma(j/2)}.
\end{equation}
As we not only disregard the ordering of the eigenvalues $z_{\pm j}$ in each $2\times 2$-block, but also the ordering of the $2\times 2$-blocks themselves, all of this has to be supplemented with a factor $1/\widetilde{N}!$.

Finally bringing everything together we obtain the correctly normalized joint eigenvalue density
\begin{equation}	\label{A1jpdfRes}
\begin{split}
G^{(1,N)}_{\mu,\nu}(z) &= C_{\mu,\nu}^{(N)} \Delta_N(z) \prod_{j=1}^{\widetilde{N}} g_{\mu,\nu}^{(1,N)}(z_{2j-1}, z_{2j}),
\\
G^{(1,N)}_{\mu,\nu}(z) &= C_{\mu,\nu}^{(N)} \Delta_N(z) h_{\mu,\nu}^{(N)}(z_N) \prod_{j=1}^{\widetilde{N}} g_{\mu,\nu}^{(1,N)}(z_{2j-1}, z_{2j})
\end{split}
\end{equation}
for $N$ even resp. odd with the antisymmetric function
\begin{equation}	\label{A1gDef}
\begin{split}
g_{\mu,\nu}^{(1,N)}(z_1, z_2) =& \left(z_1 z_2\right)^{2\nu} \frac{\abs{z_2-z_1}}{z_2-z_1}
\\
&\times \frac{B(1/2,N/2+\mu+\nu+1/2) \delta(y_1) \delta(y_2) +2\, \delta(x_1-x_2) \delta(y_1+y_2) Q_{\mu,\nu}^{(N)}(z_1, z_1^*)}{ \left[(1 + z_1^2)(1 + z_2^2)\right]^{N/2+\mu+\nu+1/2}}.
\end{split}
\end{equation}
The antisymmetry of this function is due to us disregarding an ordering \eqref{A1EVOrdering}. The normalization is
\begin{equation}
\begin{split}
C_{\mu,\nu}^{(N)} =& \frac{2^{N-3\widetilde{N}} \pi^{-\widetilde{N}^2 +\widetilde{N}(N-1)-N(N-1)/4}}{(1+N-2\widetilde{N}) \widetilde{N}!}
\prod_{j=1}^N \frac{\Gamma (\mu + \nu +(N+j)/2)}{\Gamma (\nu + j/2) \Gamma(\mu + j/2)}
\\
&\times \prod_{j=1}^{\widetilde{N}} \frac{\Gamma\left( N/2+\mu+\nu +1/2 \right) \Gamma\left( N/2+\mu+\nu+1 \right)}{\Gamma\left( N+\mu+\nu+1/2-j \right) \Gamma\left( N+\mu+\nu+1-j \right)}.
\end{split}
\end{equation}
Setting $\mu = \nu = 0$ one immediately finds the joint probability density of eigenvalues \eqref{4jpdf} of the ordinary real spherical ensemble.

The Pfaffian of the moment matrix \eqref{4MomentMatDef} can be related to this constant. One finds for the even case \cite{SommersWieczorek2008}
\begin{equation}
\int\limits_{\mathbb{C}^{2M}} d[z] \Delta_{2M}(z) \prod_{j=1}^M g^{(1,N)}(z_{2j-1},z_{2j}) = M!\, \Pf\, D^{(2M)},
\end{equation}
where $M\leq N/2$. On the other hand by equations \eqref{A1jpdfRes} and \eqref{A1gDef} this integral is
\begin{equation}
\int\limits_{\mathbb{C}^{2M}} d[z] \Delta_{2M}(z) \prod_{j=1}^M g_{(N-2M)/2,0}^{(1,2M)}(z_{2j-1},z_{2j}) = \frac{1}{C^{(2M)}_{(N-2M)/2,0}}
\end{equation}
and therefore
\begin{equation}
\Pf\, D^{(2M)} = \frac{1}{M!} \frac{1}{C^{(2M)}_{(N-2M)/2,0}}.
\end{equation}
We use this result at multiple points in the body of the text to evaluate the prefactors of our integrals.

\section{Characteristic polynomials of the real induced spherical ensemble}	\label{A2}
We are interested in integrals of the type
\begin{equation}
I_M = \int\limits_{\mathbb{C}^{2M}} d[z] \Delta_{2M}(z) \prod_{j=1}^{M} g^{(1,N)}(z_{2j-1},z_{2j}) \prod_{j=1}^{2M} (x-z_j),
\end{equation}
where $M \leq N/2$. Applying equations \eqref{A1jpdfRes} we may map them to integrals over a matrix density
\begin{equation}	\label{A2int}
\begin{split}
I_M = &\int\limits_{\mathbb{C}^{2M}} d[z] \Delta_{2M}(z) \prod_{j=1}^{M} g_{(N-2M)/2,0}^{(1,2M)}(z_{2j-1},z_{2j}) \prod_{j=1}^{2M} (x-z_j)
\\
= &\frac{1}{C^{(2M)}_{(N-2M)/2,0}} \int d[Y] \widetilde{G}_{(N-2M)/2,0}^{(1,2M)}(Y) \det(x-Y).
\end{split}
\end{equation}
The matrix density $\widetilde{G}^{(1,N)}_{\mu,\nu}(Y)$ is invariant under left and right actions of the orthogonal group, see \eqref{A1mpdf},
\begin{equation}
Y \to O_1 YO_2
\qquad
\text{with }O_1, O_2 \in \text{O}(N).
\end{equation}
Let us choose $O_1 = \mathds{1}_N$ and $O_2$ and as diagonal with entries $\pm 1$. This effectively changes the sign of $Y$ column-wise, $Y_{jl} \to \pm Y_{jl}$ for all $l$. Expanding the determinant we see that only one term survives the average
\begin{equation}
\begin{split}
\left\langle \det\left( x- Y\right) \right\rangle = \sum_{\sigma \in \mathbb{S}_N} \sgn\, \sigma \left\langle \prod_{j=1}^N \left(x \delta_{j\sigma(j)} - Y_{j\sigma(j)}\right)\right\rangle = \sum_{\sigma \in \mathbb{S}_N} \sgn\, \sigma\, x^N \prod_{j=1}^N \delta_{j\sigma(j)} = x^N,
\end{split}
\end{equation}
which is a monomial of power $N$. Therefore we obtain
\begin{equation}
I_M = \frac{x^{2M}}{C^{(2M)}_{(N-2M)/2,0}}.
\end{equation}
This result corresponds to the skew-orthogonal polynomial of even degree \cite{HKGG2023, AKP2010}.

\section{Alternative expression of \texorpdfstring{${\rm K}_3^{(1)}$}{}}

Unlike in \eqref{4Xi3Reduced} we expand the determinant only in the last column
\begin{equation}
\Xi_3 = \frac{2\, C^{(N)}}{\kappa_2-\kappa_1} \int\limits_{\mathbb{C}^{N+2}} d[z] \frac{1}{z_{N+2} + \kappa_2} \prod_{j=1}^{N/2+1} g^{(1,N)}(z_{2j-1},z_{2j}) 
\det\left[\begin{array}{c|c}
\displaystyle z_a^{b-1} & \displaystyle \frac{1}{z_a+\kappa_1}
\end{array}\right]_{\substack{1\leq a\leq N+1 \\ 1\leq b\leq N}}
\end{equation}
and apply the identity (see \eqref{4BerDetIdentity})
\begin{equation}
\begin{split}
\det\left[\begin{array}{c|c}
\displaystyle z_a^{b-1} & \displaystyle \frac{1}{z_a+\kappa_1}
\end{array}\right]_{\substack{1\leq a\leq N+1 \\ 1\leq b\leq N}} &= (-1)^{N+2} \Delta_{N+1}(z) \prod_{j=1}^{N+1} \frac{1}{z_j + \kappa_1} 
\\
&= (-1)^{N+2} \frac{\Delta_{N}(z)}{z_{N+1}+\kappa_1} \prod_{j=1}^N \frac{z_{N+1}-z_j}{z_j + \kappa_1}.
\end{split}
\end{equation}
This allows us to identify the integral over $z_1, \ldots, z_N$ as the function $\Xi_2$, see Eq.~\eqref{4Xi2}, that we calculated for the second kernel, resulting in
\begin{equation}
\begin{split}
\Xi_3 &= \frac{2\, C^{(N)}}{\kappa_2-\kappa_1}
\int\limits_{\mathbb{C}^{N+2}} d[z] \frac{\Delta_{N}(z)}{(z_{N+1}+\kappa_1) (z_{N+2}+\kappa_2)} \prod_{j=1}^N \frac{z_{N+1}-z_j}{z_j + \kappa_1} \prod_{j=1}^{N/2+1} g^{(1,N)}(z_{2j-1},z_{2j})
\\
&= \frac{2}{\kappa_2-\kappa_1} \int\limits_{\mathbb{C}^2} d[z] \frac{g^{(1,N)}(z_1,z_2)}{(z_1+\kappa_1) (z_2+\kappa_2)} \left\langle \frac{\det\left( -z_1 K_1+ K_2 \right)}{ \det\left( \kappa_1 K_1+ K_2 \right)}\right\rangle.
\end{split}
\end{equation}
Inserting our result \eqref{4Xi2Res} for the ensemble average we find the following expression for the third kernel
\begin{equation}
\begin{split}
{\rm K}_3^{(1)}(q_m,q_n) =& \frac{2}{[b(q_m) b(q_n)]^{N-1}} \int\limits_{\mathbb{C}^2} d[z] \frac{g^{(1,N)}(z_1,z_2)}{(a(q_m) + b(q_m)z_1) (a(q_n) + b(q_n)z_2)} \left\langle \frac{\det( -z_1 K_1+ K_2 )}{ \det( \kappa(q_m) K_1+ K_2 )}\right\rangle
\\
=& \frac{-N(N-1)}{2\pi [b(q_m) b(q_n)]^{N-1}} \int\limits_{\mathbb{C}^2} d[z] \frac{g^{(1,N)}(z_1,z_2)}{(a(q_m) + b(q_m)z_1) (a(q_n) + b(q_n)z_2)}
\\
&\times \Bigg[ \frac{(-1)^{N/2} 2\pi B(1/2,(N+1)/2)}{N-1} \binom{(N-1)/2}{N} 
\\
&\times { _2F_1} \left( 1,(N+1)/2;N+1;1+\left(\frac{\kappa(q_m)-z_1}{\kappa(q_m)+z_1} \right)^2 \right)
\\
&+2i \int\limits_{\mathbb{C}} d[z] \frac{z^{N-2} \sgn(\Im\, z) Q(z,z^*)}{\abs{1+z^2}^{N+1}} \left( z^* + \frac{\kappa(q_m)-z_1}{\kappa(q_m)+z_1} \right)^{-1}
\Bigg].
\end{split}
\end{equation}
Furthermore inserting \eqref{3gDef} yields
\begin{equation}
\begin{split}
{\rm K}_3^{(1)}(q_m,q_n) =& \frac{-N(N-1) b(q_m)}{2\pi b^{N-1}(q_n)} 
\Bigg(
\int\limits_\mathbb{R} dx \frac{r(x,v(q_n))}{a(q_m)+b(q_m) x} \left( \frac{x^2+1}{a(q_m)+b(q_m)x} \right)^N
\\
&\times \Bigg[\frac{(-1)^{N/2} 2\pi B(1/2,(N+1)/2)}{N-1} \binom{(N-1)/2}{N}
\\
&\times { _2F_1} \left( 1,(N+1)/2;N+1;1+\left(\frac{a(q_m)-b(q_m)x}{a(q_m)+b(q_m)x} \right)^2 \right)
\\
&+ \int\limits_\mathbb{C} d[z] s\left( z,z^*, \left(\frac{a(q_m)-b(q_m)x}{a(q_m)+b(q_m)x} \right) \right) z^{N-2}\Bigg]
\\
&+ \int\limits_\mathbb{C} d[z_1] \frac{s( z_1,z_1^*,v(q_n))}{a(q_m)+b(q_m)z_1} \left( \frac{z_1^2+1}{a(q_m)+b(q_m)z_1} \right)^N
\\
&\times \Bigg[\frac{(-1)^{N/2} 2\pi B(1/2,(N+1)/2)}{N-1} \binom{(N-1)/2}{N}
\\
&\times { _2F_1} \left( 1,(N+1)/2;N+1;1+\left(\frac{a(q_m)-b(q_m)z_1}{a(q_m)+b(q_m)z_1} \right)^2 \right)
\\
&+ \int\limits_\mathbb{C} d[z_2] s\left( z_2,z_2^*, \left(\frac{a(q_m)-b(q_m)z_1}{a(q_m)+b(q_m)z_1} \right) \right) z_2^{N-2}\Bigg] \Bigg)
\end{split}
\end{equation}
with the functions \eqref{3rsfunctions}. In this representation also a dependence solely in O(2)-invariants cannot be identified.

\bibliography{lit}

\end{document}